\documentclass[pra,twocolumn,showpacs,floatfix,superscriptaddress,aps]{revtex4}
\usepackage{amsmath}
\usepackage{makeidx}
\usepackage{amssymb}
\usepackage{graphicx}

\usepackage[usenames]{color}
\usepackage[normalem]{ulem}

\newcommand{\beq}{\begin{equation}}
\newcommand{\eeq}{\end{equation}} 
\newcommand{\beqa}{\begin{eqnarray}}
\newcommand{\eeqa}{\end{eqnarray}} 
\begin{document}

%\begin{multicols}{2}

\title{Stable, mobile, dark-in-bright, dipolar Bose-Einstein condensate
 soliton}

\author{ S. K. Adhikari\footnote{adhikari@ift.unesp.br; URL: http://www.ift.unesp.br/users/adhikari}
} 
\affiliation{
Instituto de F\'{\i}sica Te\'orica, UNESP - Universidade Estadual Paulista, 01.140-070 S\~ao Paulo, S\~ao Paulo, Brazil
} 

\begin{abstract}

We demonstrate 
robust, stable, mobile,  quasi-one-dimensional, dark-in-bright 
dipolar Bose-Einstein condensate (BEC) soliton with a notch in the central plane    
formed due to  
dipolar interaction  for repulsive  
contact interaction. At medium  velocity the head on collision of two such solitons is found to be quasi  elastic with practically no deformation.  A proposal for creating dipolar dark-in-bright solitons in 
laboratories by phase imprinting  
is also discussed. A rich variety of such solitons can be formed in dipolar 
binary BEC, where one can have a dark-in-bright soliton coupled to a bright soliton or two coupled dark-in-bright solitons. 
 The findings are illustrated using numerical simulation in three spatial  dimensions employing realistic   interaction parameters for a dipolar $^{164}$Dy BEC and a binary $^{164}$Dy-$^{162}$Dy BEC.

\end{abstract}


\maketitle

\section{Introduction}
 
A bright soliton is a self-reinforcing solitary wave in the form 
of a local peak in density
that maintains its shape, while
traveling at a constant velocity in one dimension (1D), due to a cancellation of nonlinear attraction and dispersive
effects. A dark soliton corresponds to a dip in  uniform density in 1D, which also can move with a constant velocity maintaining its shape. Solitons   have been studied in water wave, nonlinear optics, Bose-Einstein condensates (BEC) among others \cite{rmp}.  In physical three-dimensional (3D) world quasi-solitons are observed where a reduced (integrated) 1D
density exhibit soliton-like property.  
Experimentally, bright matter-wave solitons and soliton trains were created in a BEC of
$^7$Li \cite{1}
 and
$^{85}$Rb atoms \cite{3} by turning the atomic
interaction attractive from repulsive using a Feshbach resonance   \cite{fesh} and releasing
the BEC in an axially free or an expulsive trap \cite{4}. 
However, due to collapse instability, in 3D,
bright solitons are fragile and can accommodate only a small 
number of atoms.

A dark soliton corresponds to a notch (zero) in uniform 1D density,
which can propagate with a constant velocity. However, 
this condition  cannot be realized in a trapped 3D BEC, where a notch in a plane passing through the center     has been termed a dark soliton.  Such a dark soliton in a trapped BEC
 has been observed experimentally  and its small (axial) oscillation around the center has been studied \cite{darksol,dsol2,dsol}. 
However, long-time dynamics of BEC dark solitons has always been found to be unstable \cite{anglin,va,dsol} except for very {  strong transverse} trapping condition leading to a quasi-1D situation. For moderate  {  to weak transverse} traps, both theoretical and experimental considerations reveal that these dark solitons in BEC are unstable, exhibit snake instability \cite{dsol}
and eventually decay forming a vortex ring \cite{vortexring}. 
{ In addition, the trapped dark solitons can decay by a slow viscous acceleration due to their negative effective mass \cite{anglin}.}
Although, experimentally 
realizable,  dark solitons in a trapped BEC can hardly be termed a soliton, as neither the notch nor the trapped BEC can move with a constant velocity without change of shape as in the case of an integrable 1D dark or bright soliton. { Moreover, the dark solitons of a trapped BEC are realized in a fully repulsive set up, and the trap in the axial direction cannot be removed, 
as in a bright soliton, to make the dark soliton mobile in the axial direction.}

{ 
The recent study  of  BECs of $^{164}$Dy \cite{ExpDy,dy}, $^{168}$Er \cite{ExpEr} and 
  $^{52}$Cr \cite{cr,saddle}  atoms
with large magnetic dipole
moments has initiated  new investigations  of  BEC solitons in a different scenario.
It is possible to have dipolar BEC solitons for fully repulsive 
contact interaction \cite{1D}. The dipolar BEC solitons of a large numer of atoms, 
stabilized  by long-range dipolar  attraction, could be  robust and less vulnerable to collapse in 3D  due to the short-range 
contact repulsion. Quasi-1D \cite{1D}, quasi-two-dimensional (quasi-2D) \cite{2D}, vortex
\cite{ol2D} and dark \cite{dipdrk} solitons have been predicted in dipolar BEC.
 Dipolar BEC solitons can also be stabilized in  periodic optical-lattice trap(s)
replacing the usual  
harmonic trap(s)
 in quasi-1D \cite{ol1D} and quasi-2D \cite{ol2D} set-ups.}

Taking advantage of the robust nature of the large dipolar bright solitons, we consider a new class of bright solitons with a notch in 
the central radial plane and capable of moving in the axial direction 
with a constant velocity
without deformation.  We call these objects dark-in-bright solitons, which are stretched in the axial direction compared to the  bright soliton without a notch. They are stable and  stationary excitations
of the bright soliton.  
The head-on collision between two dark-in-bright solitons or between a dark-in-bright and a bright soliton is found to be
quasi  elastic at medium  velocities of few mm/s. 
In such a collision, two solitons  pass through each other without significant deformation.   { However,  as the velocity is further lowered, the collision becomes inelastic with visible deformation of the solitons during collision. The collision of solitons can be completely elastic only in 1D
integrable systems.}

 We also consider the possibility of the creation of the dark-in-bright solitons without axial trapping
by phase imprinting \cite{darksol,phase}
over a normal bright soliton with identical parameters. 
We consider the dynamical evolution of a bright soliton where the two halves have opposite phases. Upon dynamical numerical simulation such a soliton is found to develop a notch in the central radial plane between the two halves with opposite phases as in a dark soliton \cite{dsol}.  {  As the present 
dark-in-bright solitons are realized in the absence of axial trapping, unlike the conventional dark solitons of a trapped BEC, 
these are capable of moving with a constant velocity.}  
Such dark-in-bright solitons formed due to the long-range dipolar interaction are not realizable in nondipolar BECs.

These axially-free dark-in-bright solitons are so robust that they can also be realized in binary dipolar BECs. In binary BECs two stable configurations were considered: (1) one distinct dark-in-bright soliton in each component and (2) a dark-in-bright soliton in one component coupled to a bright soliton in the other component.

{ In Sec. II the time-dependent 3D mean-field model
for the binary dipolar BEC soliton is presented. The
results of numerical calculation are exibited in Sec. III.
The domain of a stable bright and dark-in-bright  solitons  is illustrated in
stability phase diagram showing the maximum number of  $^{164}$Dy and $^{168}$Er
 atoms versus respective scattering lengths. The dynamical evolution of collision
between  two dark-in-bright  solitons and between a bright and a  dark-in-bright  soliton
is considered. The evolution of a phase imprinted bright soliton  to a dark-in-bright soliton
is also demonstrated. Stability phase diagram  for the appearance of dark-in-bright solitons in the binary
$^{164}$Dy-$^{162}$Dy mixture is also considered. 
     Finally, in Sec. IV a
brief summary of our findings is presented.

}

\section{Mean-field model}

{ The extension of the mean-field Gross-Pitaevskii (GP)
equation to a binary dipolar boson-boson \cite{mfb2} and
boson-fermion \cite{bf} mixtures are well established, and,
for the sake of completeness, we make a brief summary
of the same appropriate for this study.
}   We present the binary GP equations in dimensionless form which are more practical to use and have a neater look.

\subsection{Binary BEC}

We consider a binary dipolar BEC soliton, with the 
mass, number of atoms, magnetic  dipole moment, and scattering length for the two species $ j=1,2,$
given by $m_j, N_j, 
\mu_j, a_j,$ respectively.   The intra- ($V_{j}$)
and interspecies ($V_{12}$)
interactions 
for two atoms at  $\bf r$ and $\bf r'$ are 
\begin{eqnarray}\label{intrapot} 
V_j({\bf R})= 3
a_{\mathrm {dd}}^{(j)}V_{\mathrm {dd}}({\mathbf R})+4\pi a_j \delta({\bf R})
,\\
V_{12}({\bf R})= 3
a_{\mathrm {dd}}^{(12)}V_{\mathrm {dd}}({\mathbf R})/2+2\pi a_{12} \delta({\bf R}),
\label{interpot} 
     \end{eqnarray}
respectively, with \begin{eqnarray}\label{dipx}  a_{\mathrm {dd}}^{(j)}=
\frac{\mu_0  \mu_j^2m_j}{12\pi \hbar ^2    },\quad
a_{\mathrm {dd}}^{(12)}=
\frac{\mu_0  \mu_1 \mu_2m_1m_2}{6\pi \hbar ^2(m_1+m_2)    }, 
\\
V_{\mathrm {dd}}({\mathbf R})=\frac{1-3\cos^2 \theta}{{\bf R}^3},
\end{eqnarray}
where $a_{12}$ is the intraspecies scattering length, $\mu_0$ is the permeability of free space, 
$\theta$ is the angle made by the vector ${\bf R}$ with the polarization 
$z$ direction,  $\bf R = (r-r').$ 
{ The strengths of intra- and interspecies dipolar interactions are here expressed in terms  
 of the dipolar lengths  $a_{\mathrm {dd}}^{(j)}$ and $a_{\mathrm {dd}}^{(12)}$ given by Eq. (\ref{dipx}) 
in the same way as the strengths of contact interactions
are expressed in terms of scattering lengths $a_j$ and $a_{12}$.}
The dimensionless GP equations for the axially-free  quasi-1D binary soliton  
can be written as   \cite{mfb2}
\begin{align}& \,
 i \frac{\partial \phi_1({\bf r},t)}{\partial t}=
{\Big [}  -\frac{\nabla^2}{2 }
+
\frac{1}{2} \rho^2
\nonumber \\  &  \,
+ g_1 \vert \phi_1 \vert^2
+ g_{12} \vert \phi_2 \vert^2
+ g_{\mathrm {dd}}^{(1)}
\int V_{\mathrm {dd}}({\mathbf R})\vert\phi_1({\mathbf r'},t)
\vert^2 d{\mathbf r}' 
\nonumber \\  &  \,
+ g_{\mathrm {dd}}^{(12)}
\int V_{\mathrm {dd}}({\mathbf R})\vert\phi_2({\mathbf r'},t)
\vert^2 d{\mathbf r}' 
{\Big ]}  \phi_1({\bf r},t),
\label{eq3}
\\
%\end{align}
%\begin{align}
& \,
{ i} \frac{\partial \phi_2({\bf r},t)}{\partial t}={\Big [}  
-m_{12} \frac{\nabla^2}{2}
+
\frac{1}{2}m_\omega \rho^2
\nonumber \\  &  \,
+ g_2 \vert \phi_2 \vert^2 
+ g_{21} \vert \phi_1 \vert^2 
+ g_{\mathrm {dd}}^{(2)}
\int V_{\mathrm {dd}}({\mathbf R})\vert\phi_2({\mathbf r'},t)
\vert^2 d{\mathbf r}'
\nonumber \\ & \,
+ g_{\mathrm {dd}}^{(21)}
\int V_{\mathrm {dd}}({\mathbf R})\vert\phi_1({\mathbf r'},t)
\vert^2 d{\mathbf r}'  
{\Big ]}  \phi_2({\bf r},t),
\label{eq4}
\end{align}
where
$\rho^2=x^2+y^2,\quad  i=\sqrt{-1},$
%\begin{align}&
 $m_\omega=\omega_2^2/(m_{12}\omega_1^2),$
$m_{12}={m_1}/{m_2},$
$g_1=4\pi a_1 N_1,$
$g_2= 4\pi a_2 N_2 m_{12},$
$g_{12}={2\pi m_1} a_{12} N_2/m_R,$
$g_{21}={2\pi m_1} a_{12} N_1/m_R,$
$g_{\mathrm {dd}}^{(2)}= 3N_2 a_{\mathrm {dd}}^{(2)}m_{12},$
$g_{\mathrm {dd}}^{(1)}= 3N_1 a_{\mathrm {dd}}^{(1)},$
$g_{\mathrm {dd}}^{(12)}= 3N_2 a_{\mathrm {dd}}^{(12)}m_1/(2m_R),$
$g_{\mathrm {dd}}^{(21)}= 3N_1 a_{\mathrm {dd}}^{(12)}m_{1}/(2m_R),$
and where $\omega_j$  is the radial frequency 
of the harmonic trap acting on species $j$. 
In Eqs. (\ref{eq3}) and (\ref{eq4}), length is expressed in units of 
oscillator length  $l=\sqrt{\hbar/(m_1\omega_1)}$, 
energy in units of oscillator energy  $\hbar\omega_1$, probability density 
$|\phi_j|^2$ in units of $l^{-3}$, and time in units of $ 
t_0=1/\omega_1$. 

\subsection{Single-component BEC}

The dimensionless GP equation for a single-component 
dipolar quasi-1D soliton is \cite{1D}
\begin{align}& \,
{\mbox i} \frac{\partial \phi({\bf r},t)}{\partial t}=
{\Big [}  -\frac{\nabla^2}{2 }
+
\frac{1}{2} \rho^2+ 4\pi a N \vert \phi ({\mathbf r},t)\vert^2
\nonumber \\  &  \,
+ 3a_{\mathrm {dd}}N
\int V_{\mathrm {dd}}({\mathbf R})\vert\phi({\mathbf r'},t)
\vert^2 d{\mathbf r}'  
{\Big ]}  \phi({\bf r},t),
\label{single}
\end{align}
where $N$ is the number of atoms, $a$ is the scattering length, $a_{\mathrm {dd}}$ the dipolar length.

 \section{Numerical Results}

 The $^{164}$Dy and $^{168}$Er atoms 
have the largest magnetic moments of all the dipolar atoms used in BEC experiments. For the single-component dipolar BEC we consider $^{164}$Dy
atoms and for the binary dipolar BEC we consider the  $^{164}$Dy-$^{162}$Dy mixture. 
  The magnetic moment of a single $^{164}$Dy  or $^{162}$Dy
atom is  $ \mu_1 = 10\mu_B$
\cite{ExpDy} and of a $^{168}$Er atom is $ \mu_2 = 7\mu_B$ \cite{ExpEr}
with 
$\mu_B$ the Bohr magneton leading to the dipolar lengths $a_{\mathrm {dd}}(^{164}$Dy$) \approx 132.7a_0$, 
$a_{\mathrm {dd}}(^{168}$Er$)\approx 66.6a_0$, $a_{\mathrm {dd}}(^{162}$Dy$) \approx 131.0a_0$, 
and $a_{\mathrm {dd}}(^{164}$Dy$-^{162}$Dy$) \approx 131.9a_0$,  with $a_0$ the Bohr radius. 
The dipolar interaction in $^{164}$Dy
atoms is roughly double of that in $^{168}$Er atoms and 
about  eight times larger than that in $^{52}$Cr
atoms with a dipolar length  $a_{\mathrm {dd}}\approx 15a_0$ 
\cite{cr}. In both the single-component case and in the binary mixture
we take $l =1$ $\mu$m. In a single component $^{164}$Dy BEC this corresponds to a radial angular trap frequency $\omega =  2\pi \times 
61.6$ Hz corresponding to $t_0= 2.6$ ms and in a $^{168}$Er BEC this corresponds to
$\omega =  2\pi \times 
60.2$ Hz. In the binary $^{164}$Dy-$^{162}$Dy mixture $\omega_1=\omega_2=2\pi\times 61.6$ Hz.

\begin{figure}[!t]

\begin{center}
\includegraphics[width=\linewidth]{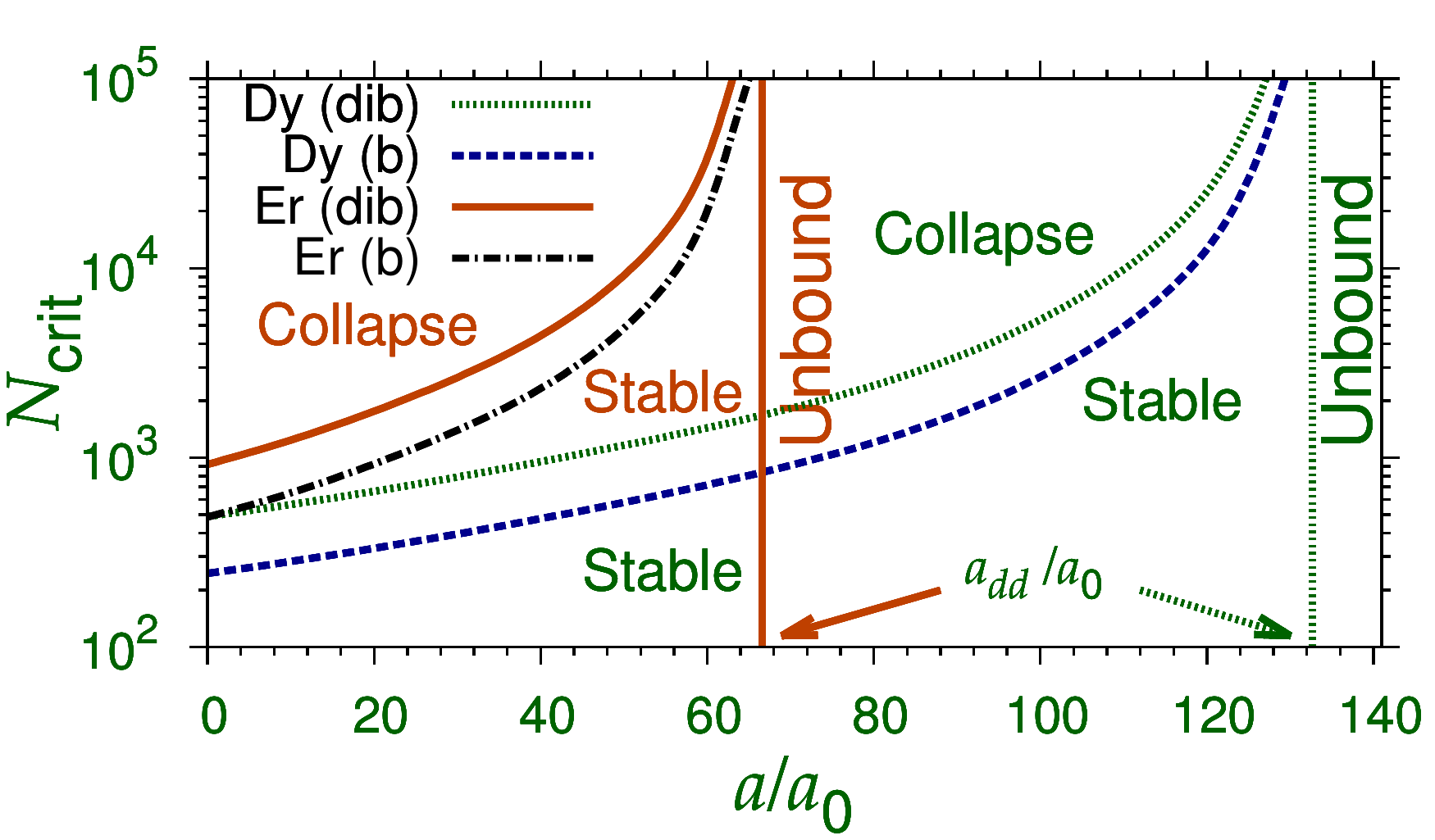}

\caption{ (Color online) Stability phase diagram showing the critical number of atoms $N_{\mathrm{crit}}$
in a  quasi-1D dipolar dark-in-bright (dib) and bright (b)
BEC soliton of $^{164}$Dy or $^{168}$Er atoms from numerical calculation. The system is repulsive and unbound for $a\gtrsim a_{\mathrm {dd}}$. Stable quasi-1D solitons appear for $a\lesssim a_{\mathrm {dd}}$ and the number of atoms $N$ below the critical number $N_{\mathrm{crit}}$.
The oscillator length
$l= 1 $ $\mu$m.
}\label{fig1} \end{center}

\end{figure}

We solve the 3D Eqs. (\ref{eq3}) and (\ref{eq4}) or Eq. (\ref{single})
by the split-step 
Crank-Nicolson discretization scheme using both real- and imaginary-time propagation
  in 3D Cartesian coordinates independent of the underlying 
trap symmetry 
using a space step of 0.1 $\sim$ 0.2
and a time step of 0.0004 $\sim$ 0.005 \cite{CPC}.  The dipolar potential term is treated by Fourier transformation  in momentum space using a convolution theorem in usual fashion \cite{Santos01}. 
It was conjectured   that stable quasi-1D dark solitons, with antisymmetric wave function, are  the stationary
lowest axial excitation   
of the system. { They are comparable to the lowest 
axial excitation of a 3D harmonic oscillator with a notch. 
The imaginary-time simulation converges to the lowest-energy solution with the specific symmetry of the initial state. For example, in the 1D linear harmonic oscillator problem, an antisymmetric initial state leads, in imaginary-time simulation, to the first excited state.}
Similarly, in  imaginary-time simulation the stationary
dark-in-bright solitons  can be obtained with an initial antisymmetric trial function, for example, 
$\phi({\bf r}) \sim z \exp[-\rho^2/2-\alpha^2 z^2/2]$  with a notch at $z=0$ and with 
a small $\alpha$  denoting large spatial axial extension of the dark-in-bright soliton.
{  The dark-in-bright
soliton is the simplest possible soliton (after the bright soliton) in a dipolar BEC.}

\begin{figure}[!t]

\begin{center}
\includegraphics[width=\linewidth,clip]{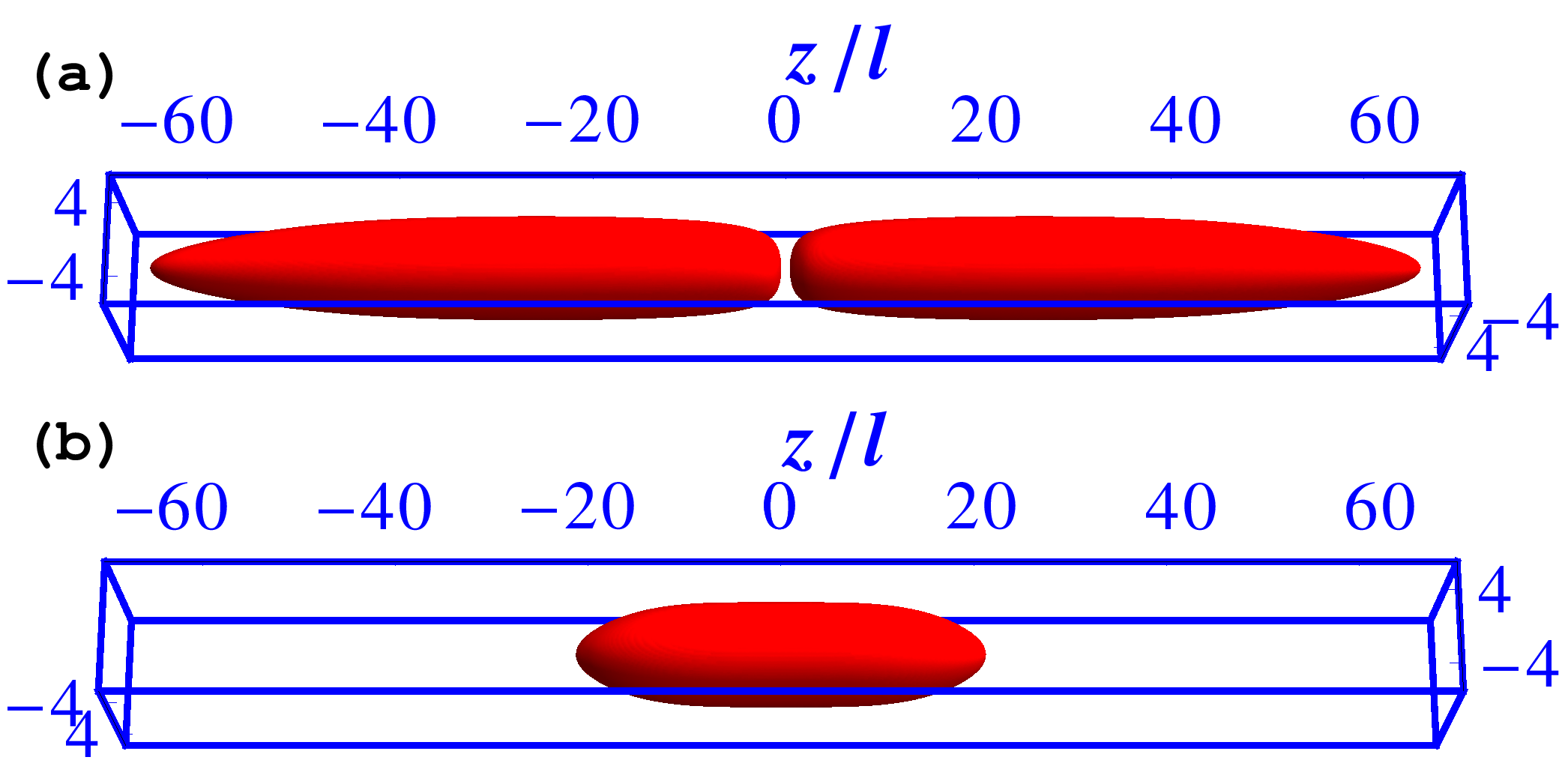}

\caption{ (Color online) 3D isodensity contour ($|\phi|^2$) of a 
(a) dark-in-bright and a (b) bright soliton of 1000 $^{164}$Dy atoms
 with  $a= 80a_0$. The dimensionless lengths $x,y,$ and $z$ are in units of $l$ $(\equiv 1$ $\mu$m).
The density on the contour is $ 10^{7}$ atoms/cm$^3$ compared to the central density in the bright soliton of (b) of about $10^{12}$ atoms/cm$^3$. 
}\label{fig2} \end{center}

\end{figure}

\begin{figure}[!t]

\begin{center}
\includegraphics[width=\linewidth,clip]{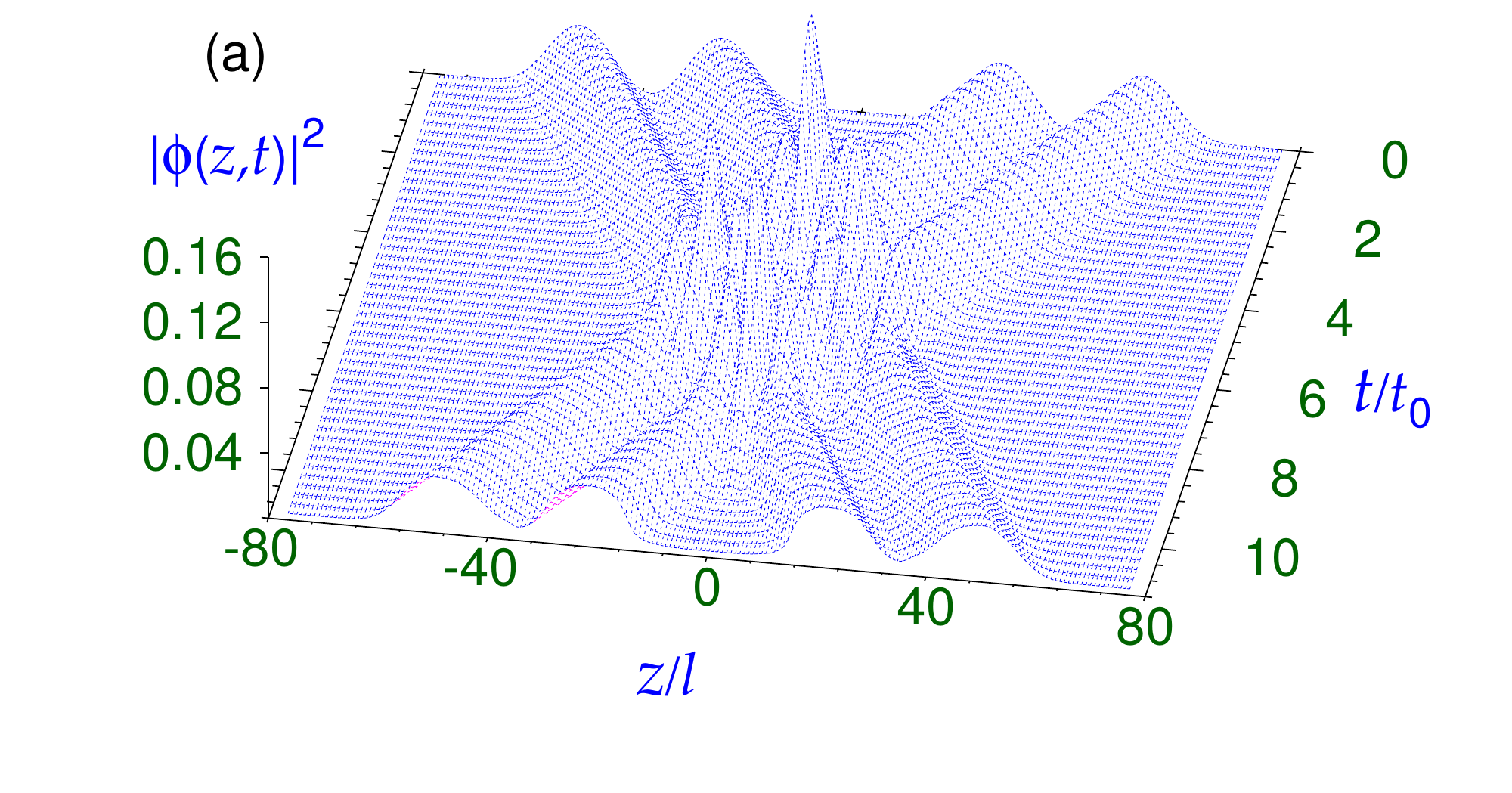}
 \includegraphics[width=\linewidth,clip]{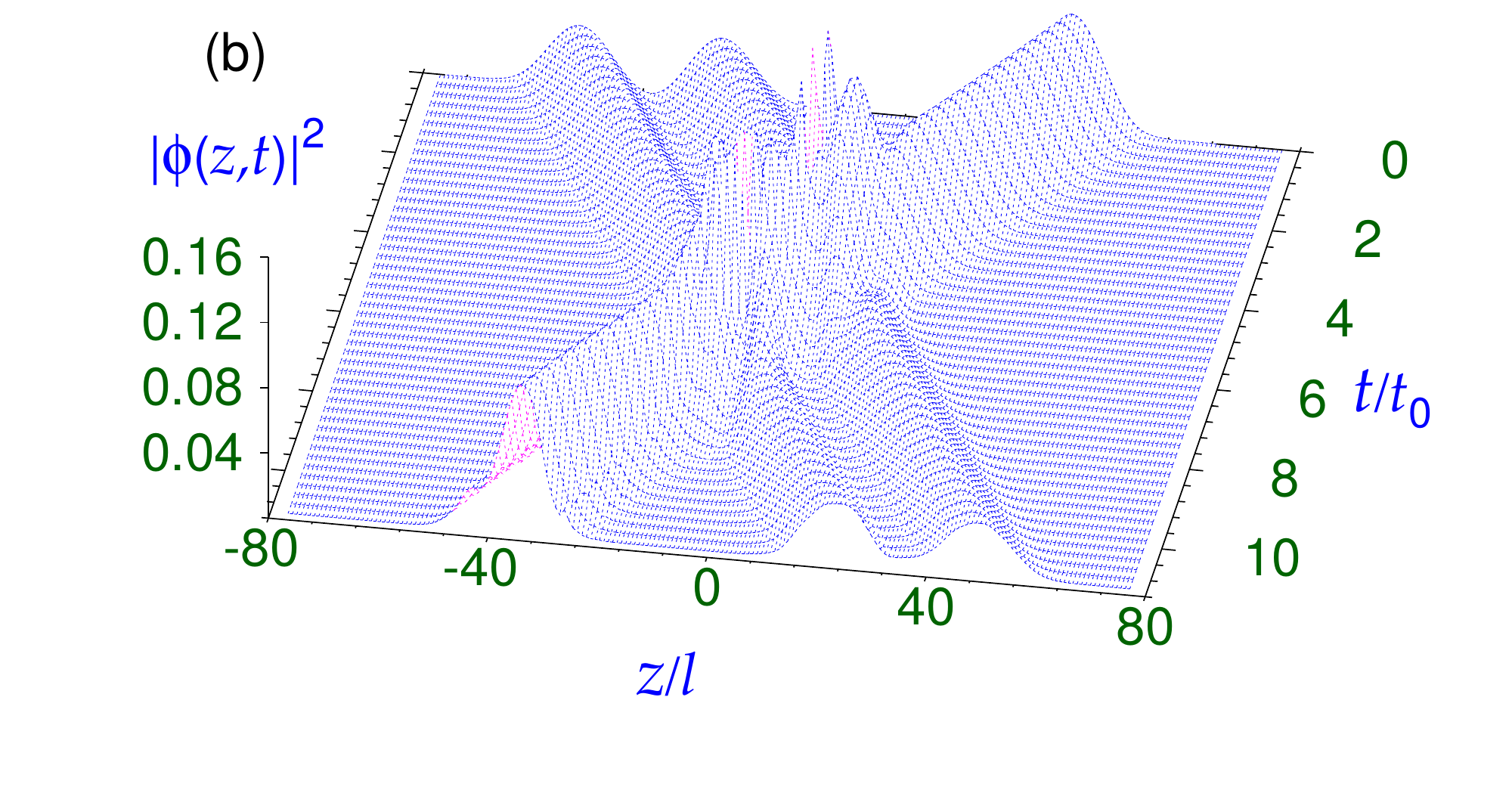}

\caption{(Color online)  
Linear axial density of (a) two colliding dark-in-bright solitons
of 1000 $^{164}$Dy atoms each
of Fig. \ref{fig2} (a) and  of (b) the colliding  dark-in-bright soliton of  Fig. \ref{fig2} (a) and a bright soliton of 500 $^{164}$Dy atoms with $a=80a_0$. 
}\label{fig3} 

\end{center}

\end{figure}

\subsection{Single-component BEC}

  We solve Eq. (\ref{single}) for different values of the scattering length $a$.
We find that for interaction parameters of $^{164}$Dy and $^{168}$Er atoms the dark-in-bright and bright solitons are stable up to a critical maximum number of atoms, beyond which the system collapses \cite{jbohn}. In Fig. \ref{fig1} we plot this critical 
number $N_{\mathrm{crit}}$ versus $a/a_0$ from numerical simulation.    
We find that a stable soliton is possible for
 $a \lesssim a_{\mathrm {dd}}$ and for a number of atoms below this critical number \cite{1D}. The critical number of atoms 
increases with the increase of contact repulsion as 
$a\to a_{\mathrm {dd}}$, which is counter-intuitive. The solitons are bound by long-range dipolar interaction and an increase of contact repulsion gives more stability against collapse for a fixed dipolar interaction strength. 
In this phase diagram three regions are shown: stable, collapse and unbound. In the unbound region ($a\gtrsim a_{\mathrm {dd}}$) contact repulsion dominates over dipolar attraction and the soliton cannot be bound. In the collapse region, the opposite happens and the soliton collapses 
due to an excess of dipolar attraction along the axial $z$ direction. In the stable region there is a balance between attraction and repulsion and a stable soliton can be formed. In Figs. \ref{fig2} (a) and (b) we show the isodensity contour of a dark-in-bright and a bright soliton of 1000 $^{164}$Dy atoms for $a=80a_0$ and $l=1$ $\mu$m. The bright soliton is much more compact with a large central density 
compared to the well-stretched
dark-in-bright soliton with a zero central density, both free to move along the axial polarization direction due to an absence of the axial trap.

Both dark-in-bright and bright solitons are unconditionally stable and 
last for ever in real-time propagation without any visible change of shape.  The dipolar attraction provides binding of the soliton and the contact repulsion reduces collapse instability. In order to have  
very robust solitons one should have $a_{\mathrm{dd}} >> a >> 0$ 
corresponding to more dipolar attraction over a sizable contact repulsion.
For $^{164}$Dy atoms with $a_{\mathrm{dd}}= 132.7a_0$, we considered $a=80a_0$ for the illustration consistent with this inequality, which can be achieved using a Feshbach resonance \cite{fesh}.

\begin{figure}[!t]

\begin{center}
\includegraphics[width=\linewidth]{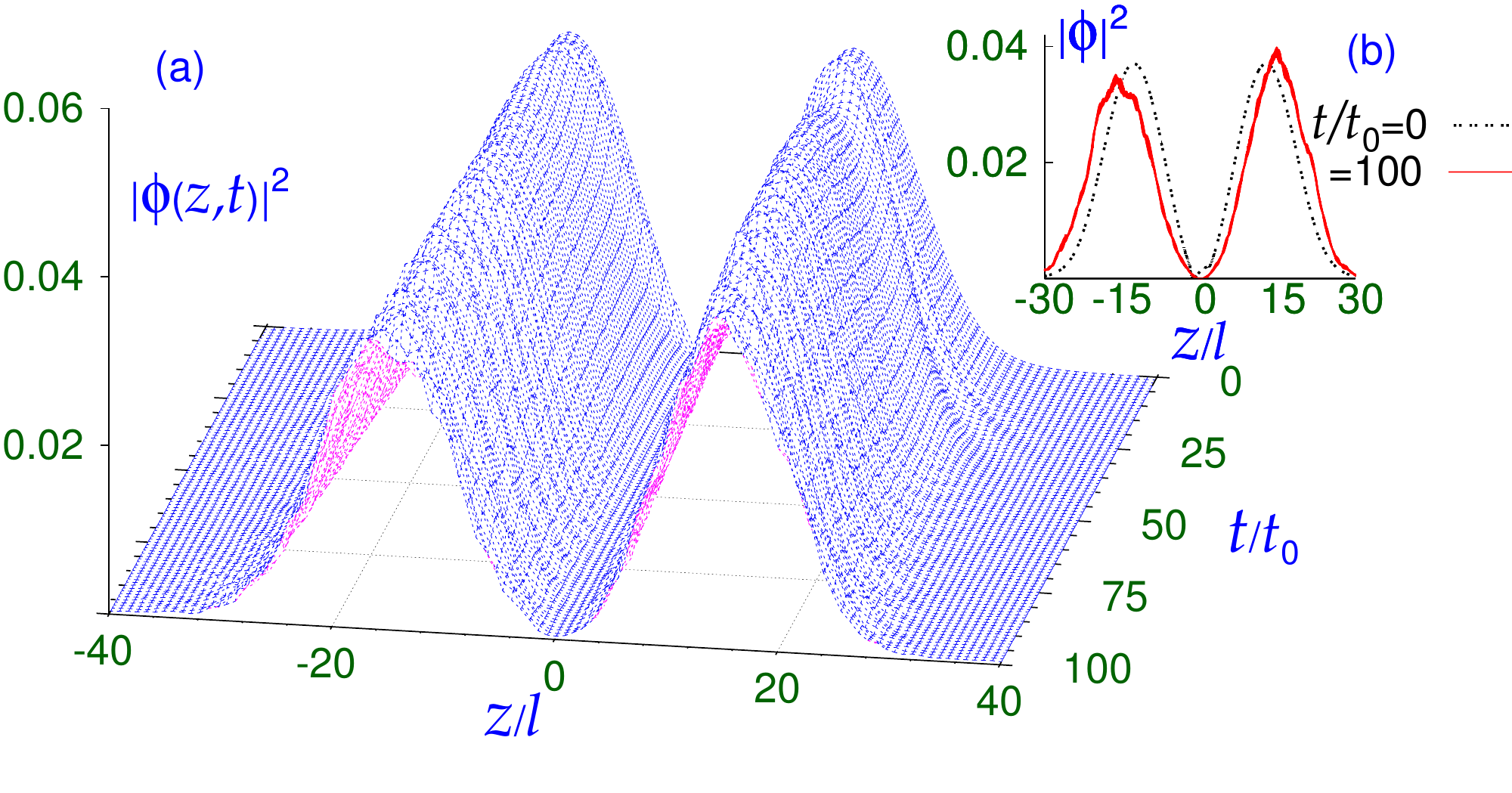}
 
{ 
\caption{ (Color online)  
(a) 
Linear axial density of the dipolar dark-in-bright soliton of 1000 $^{164}$Dy atoms with $a=80a_0$ and $l=1$ $\mu$m upon real-time propagation. The initial-state has been modified to have the central zero of the dark-in-bright soliton at $z/l \approx -2$. 
  In (b)     the initial ($t=0$) and the final ($t/t_0=100$) profiles of linear axial densities
are shown.  
}\label{fig4} 
}
\end{center}

\end{figure}

To demonstrate further the robustness of the solitons we consider a head-on collision between two solitons moving along the polarization $z$ axis in opposite directions. First, we consider the collision between two identical dark-in-bright solitons of Fig. \ref{fig2} (a) each of 1000 $^{164}$Dy atoms. 
Next we consider the collision between the dark-in-bright soliton of Fig. \ref{fig2} (a) with a bright soliton of 500 $^{164}$Dy atoms with $a=80a_0$ and $l=1$ $\mu$m.
The constant velocity of about 2.4 mm/s of each   of the colliding solitons was achieved by phase imprinting with  factors of 
$\exp(\pm \mathrm{i} 7.5 z)$ applied to the respective wave functions.
  In Fig. \ref{fig3} (a) we plot the integrated 1D density 
$|\phi(z,t)|^2=\int dx dy |\phi({\bf r},t)|^2$ of the moving dark-in-bright solitons versus 
$z$ and $t$. 
 The collision dynamics of a bright soliton with a dark-in-bright soliton
is illustrated in Fig. \ref{fig3} (b) via a plot of  the integrated 1D density of the two solitons.  After collision, the solitons emerge in both cases without a visible change of shape demonstrating the solitonic nature.  {  However, at much lower incident velocities the collision becomes inelastic and a distortion in the shape of the emerging solitons is found. This is expected, as only the collision between two 1D integrable solitons is truely   elastic.}

{ 
In case of a normal dark soliton in a trapped BEC, long-time simulation in real-time 
propagation may lead to a destruction of the dark soliton by snake instability \cite{dsol} and by instability 
against  oscillation of the central zero along the axial direction  \cite{anglin}. We tested that 
the dipolar dark-in-bright soliton maintains its profile in long-time 
real-time propagation without snake instability (not presented in this paper). This is not surprising as a dark soliton in a trapped BEC exhibits snake instability in the presence of a strong axial trap and 
does not exhibit this instability in the limit of a weak axial trap. So it is reasonable that the present dark-in-bright soliton without axial trap does not exhibit snake instability.  
Now we test the stability of the dipolar dark-in-bright soliton when the central zero  of the dark soliton is given a small displacement  with respect
to the center of the bright solitonic profile. For this test, we consider the dark-in-bright soliton of 1000 $^{164}$Dy atoms with $a=80a_0$ and modify the initial profile between $z/l=\pm 2$ and move the central zero  of the dark-in-bright soliton from $z=0$ to $z/l\approx -2$. With this modified initial profile of the dark-in-bright soliton we perform real-time simulation upto $t/t_0 =100$.  We find that the zero of the dark-in-bright soliton quickly moves to $z=0$ and no 
unstable oscillation of this zero is noted. The dark-in-bright soliton does turn to a grey-in-bright soliton with the central notch having nonzero density. 
This is illustrated by a plot of linear axial density versus time in Fig. \ref{fig4} (a). 
In Fig. \ref{fig4} (b), we show the initial and final axial densities at $t/t_0 = 0$ and 100.  
As the initial state in this study is not a stationary state, oscillation in density is noted, 
however, maintaining the central notch of the dark soliton fixed at $z=0$, confirming the stability of the dark-in-bright soliton.
 }

\begin{widetext}

\begin{figure}[!b]

\begin{center}
\includegraphics[width=\linewidth]{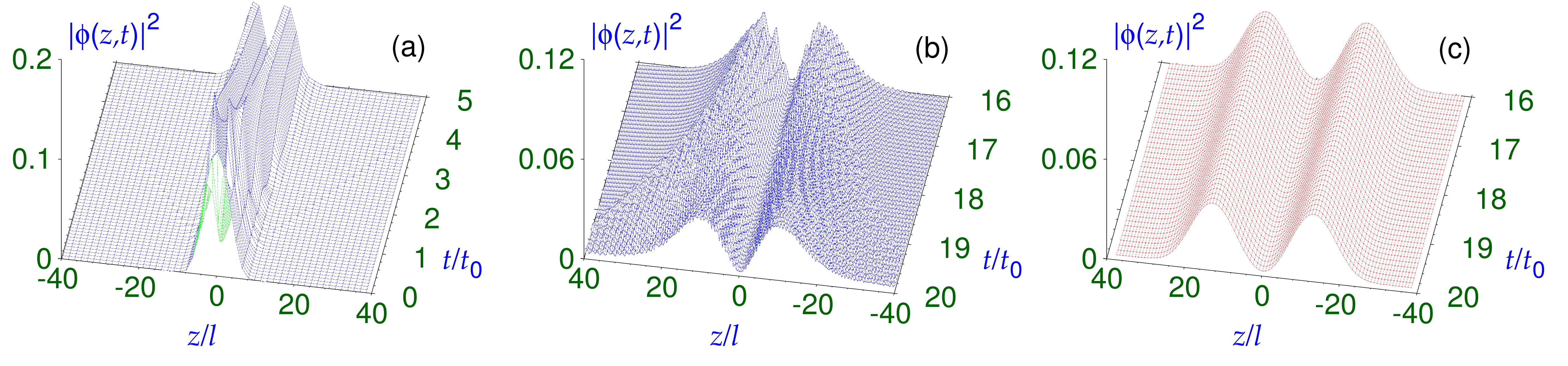}

\caption{ (Color online)  
Creating a dark-in-bright soliton on real-time simulation of a phase imprinted bright soliton of 1000 $^{164}$Dy atoms with $a=80a_0$ and $l=1$ $\mu$m.  Linear axial density of the bright soliton
(blue) for (a) small ($5>t/t_0>0$) and (b) large ($20>t/t_0>16$) times. In (c) the constant density of the stationary dark-in-bright soliton (red) of 1000 $^{164}$Dy atoms obtained by imaginary-time routine  with $a=80a_0$ and $l=1$ $\mu$m
is also shown for a comparison.    
}\label{fig5} \end{center}

\end{figure}

\end{widetext}

As the dark-in-bright solitons are stable and robust, they can be prepared by phase imprinting \cite{phase} a bright soliton.
In experiment a homogeneous potential generated using  a far detuned 
laser beam is applied on one half of the bright soliton ($z<0$) for an interval of time so as to imprint an extra phase of $\pi$ on the wave function for $z<0$ \cite{darksol}. The thus phase-imprinted  bright soliton is propagated in real time, while it slowly transforms into a dark-in-bright soliton. The present simulation is done with no axial trap. In actual experiment a very weak axial trap can be kept during generating the dark-in-bright soliton and eventually removed.  The simulation is illustrated in Figs. \ref{fig5} (a) and (b), where we plot the linear axial density of the phase-imprinted soliton versus time at small and large times. It is demonstrated that at large times the linear density tends towards that of the stable       dark-in-bright soliton with a prominent notch at the center presented in Fig. \ref{fig5} (c). 

\subsection{Binary-BEC}

\begin{figure}[!b]

\begin{center}
\includegraphics[width=\linewidth]{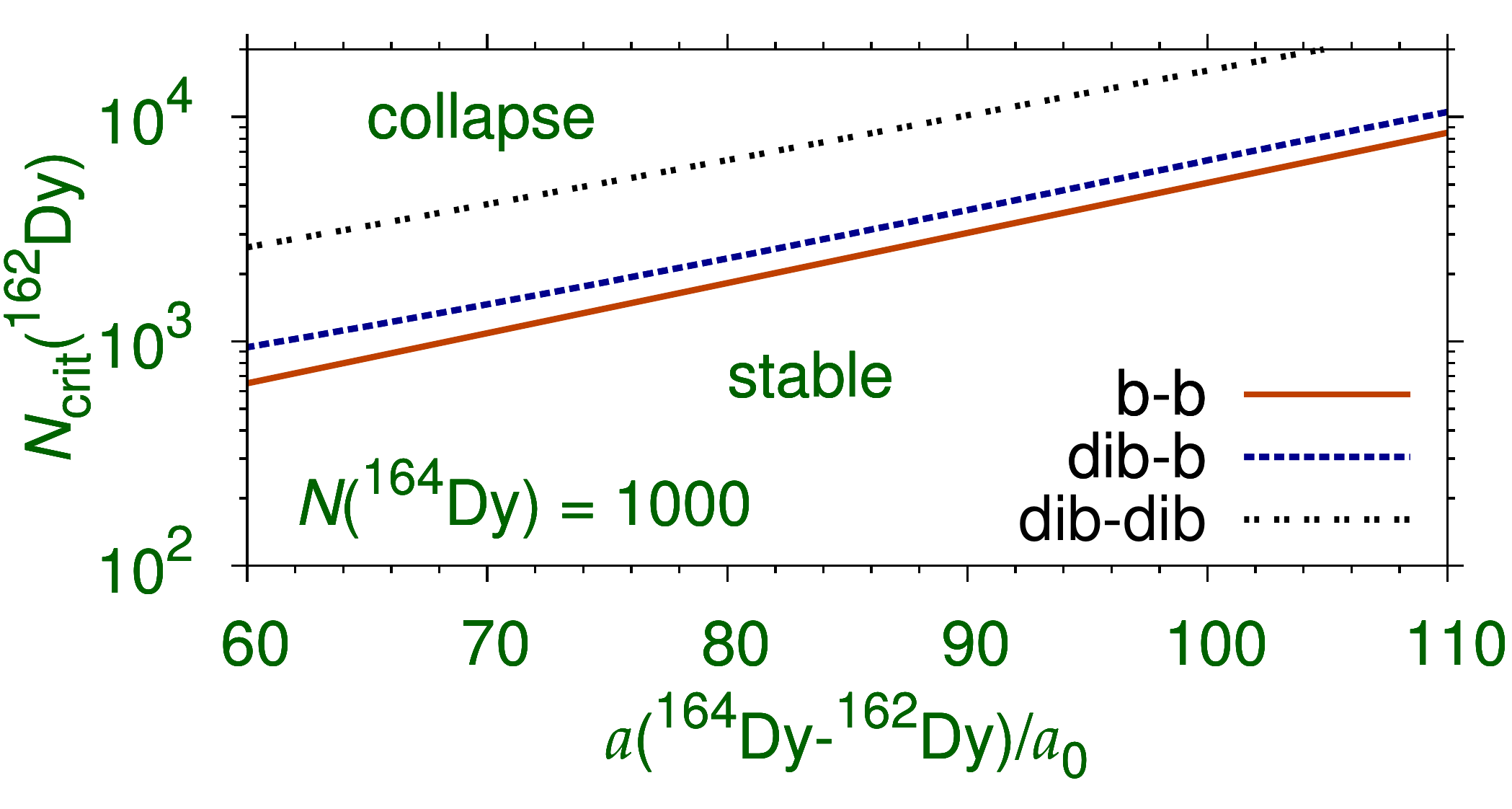}

\caption{ (Color online)  
 Stability phase diagram for the critical number of $^{162}$Dy
atoms in a binary $^{164}$Dy-$^{162}$Dy soliton with 1000 $^{164}$Dy
atoms versus the interspecies scattering length 
with (a) two coupled bright (b-b) solitons, (b) a dark-in-bright (dib) 
$^{164}$Dy soliton coupled to a bright (dib-b) $^{162}$Dy soliton, and (c) 
two coupled dib (dib-dib) solitons. Parameters used: $a(^{164}$Dy) = $a(^{162}$Dy) = $120a_0,  l=1$ $\mu$m. 
}\label{fig6} \end{center}

\end{figure}

The dark-in-bright solitons can also be realized in a binary dipolar BEC \cite{mfb}. For an illustration we consider  the $^{164}$Dy-$^{162}$Dy mixture. This is particularly interesting as Lev and his collaborators are studying this binary mixture in laboratory at the Stanford University \cite{levb}. In order to permit a large number of atoms in the solitons we consider a large value for the scattering lengths, e.g.
$a(^{162}$Dy) = $a(^{164}$Dy) =120$a_0$, viz. Fig. \ref{fig1}. The interspecies scattering length is considered as a variable. There could be two types of new binary solitons bound by interspecies attraction \cite{solmol}: (a) two coupled dark-in-bright solitons and (b) a dark-in-bright soliton coupled to a bright soliton.  
First we consider the stability phase plot for these two cases. In Fig. \ref{fig6} we show the maximum critical number of $^{162}$Dy atoms in the stable binary soliton with 1000 $^{164}$Dy atoms. 
As the mass and dipolar lengths are almost the same for the two isotopes, the binary plot is quasi symmetric under an exchange of 
$ ^{162}$Dy and $ ^{164}$Dy atoms.  The plot for   1000 $^{164}$Dy atoms in the binary soliton will be practically the same as that in Fig. \ref{fig6} with the role of the two isotopes interchanged.  As a dark-in-bright soliton can accommodate more atoms than a bright soliton, two coupled dark-in-bright solitons 
can have more atoms than a dark-in-bright soliton coupled to a bright soliton.

\begin{figure}[!t]

\begin{center}
\includegraphics[width=.9\linewidth,clip]{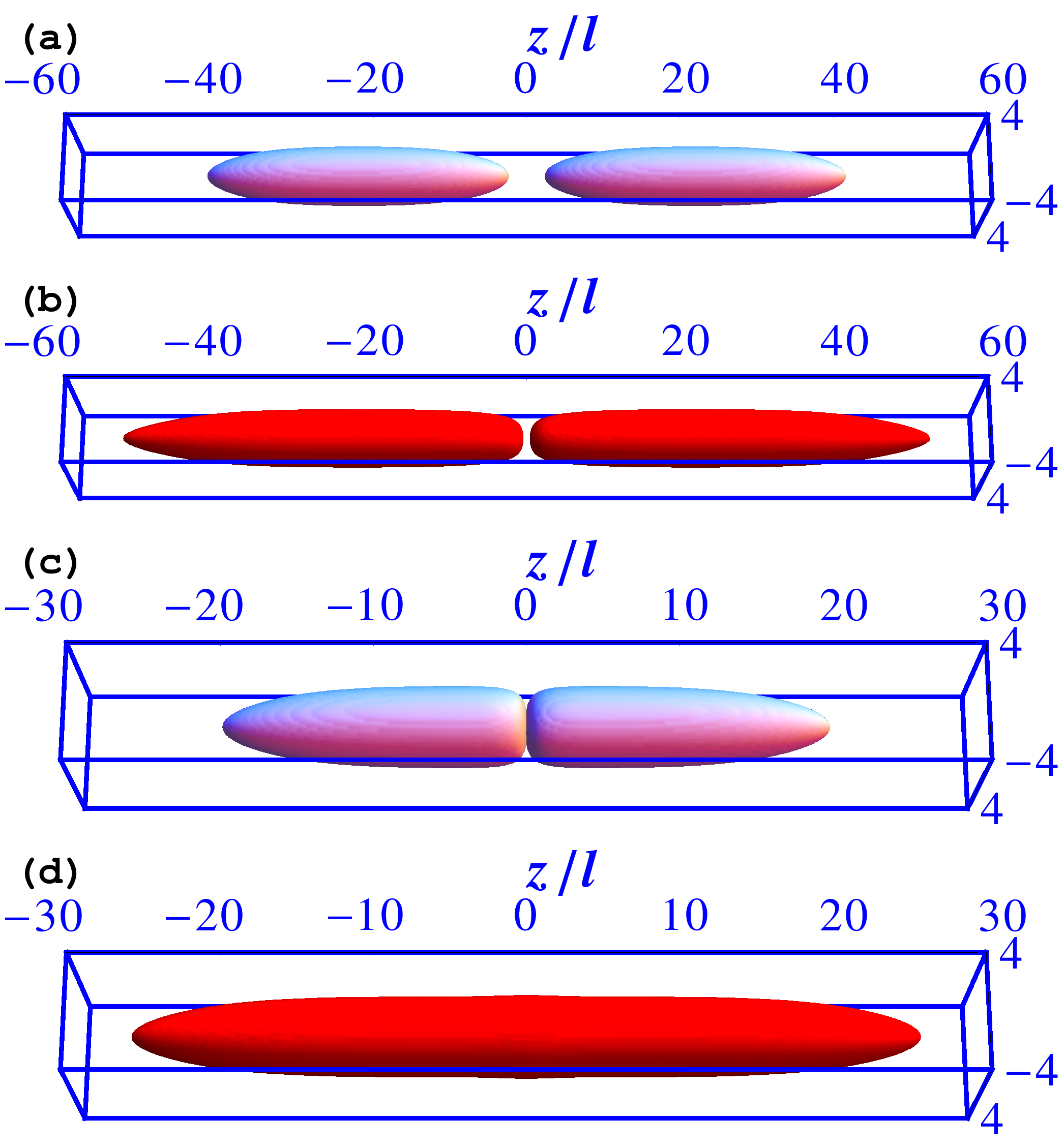}

\caption{ (Color online) 3D isodensity contour $(|\phi_i({\bf r})|^2)$ of two   dark-in-bright solitons in the binary $^{164}$Dy-$^{162}$Dy
mixture: (a)  $^{164}$Dy, (b) $^{162}$Dy profiles.
The same of a   dark-in-bright and a bright soliton in the binary $^{164}$Dy-$^{162}$Dy
mixture: (c)  $^{164}$Dy, (d) $^{162}$Dy profiles.   
Parameters used: $a$($^{164}$Dy-$^{162}$Dy)$ =100a_0$, 
$a$($^{164}$Dy)$ =$
$a$($^{162}$Dy)$ =120a_0$,
 $l = 1$ $\mu$m,
$N(^{164}$Dy) =1000, $N(^{162}$Dy) =3000. The density on the contour is  $10^{7}$ atoms/cm$^3$.
}\label{fig7} \end{center}

\end{figure}

In Fig. \ref{fig7} we show the isodensity contour of a   
binary $^{164}$Dy-$^{162}$Dy soliton for 1000  $^{164}$Dy atoms and 3000 
$^{162}$Dy atoms for the interspecies scattering length 
 $a(^{164}$Dy-$^{162}$Dy)=$100a_0$ and  intraspecies scattering lengths  $a(^{164}$Dy)=$a(^{162}$Dy)=$120a_0$.
In Figs. \ref{fig7} (a) and (b) we show the 
profiles of the coupled dark-in bright solitons of  $^{164}$Dy
and $^{162}$Dy atoms, respectively. In Figs. \ref{fig7} (c) and (d) we illustrate the profiles of the  dark-in bright $^{164}$Dy soliton 
coupled to the bright $^{162}$Dy soliton.  
 The component $^{164}$Dy  with a 
smaller number of atoms has a smaller spatial extension  whereas the component  $^{162}$Dy with a larger number of atoms has a larger spatial extension.  As the bright soliton with a large central density 
has a smaller spatial extension compared to a dark-in-bright soliton, 
the spatial extension of the bright soliton in Fig. \ref{fig7} (d) is much smaller than the bright-in-dark soliton in Fig. \ref{fig7} (b) (note the different length scales in these plots).
These binary dark-in-bright solitons are found to be stable in real-time propagation 
upon small perturbation.

%To demonstrate the stability, we performed real-time simulation with the binary coupled dark-in-bright and bright solitons of Figs. \ref{fig7} (c) and (d). To start the dynamics at $t=0$ we put the two components at  
%$z/l= \pm 6$, and attributed  velocities to the components towards $z=0$  by imprinting with phases $\exp(\pm 7.5\mathrm{i}z)$. The long-time dynamics of binary solitons is illustrated in Fig. \ref{fig7} where we show the stable small oscillations of the binary soliton around $z=0$ by the contour plot of the 1D sendity $|\phi(z,t)|^2$ versus $t/t_0$ and $z/l$. 

\section{Conclusion}

We demonstrated  the possibility of creating mobile, stable,  quasi-1D, dark-in-bright solitons in dipolar BEC with a notch in the central plane  capable of moving along the axial polarization direction with a constant velocity.  {  The snake instability in trapped BEC dark solitons  exists only for a weak transverse trap and disappears for a strong transverse trap \cite{dsol,anglin,va}.   
}
{ The present solitons are stationary solutions of the mean-field GP equation and being axially free with a strong transverse trap
they do not exhibit snake instability.} 
The head on collision between two dark-in-bright solitons or between a bright and a dark-in-bright soliton with a relative velocity of about 
5 mm/s is quasi elastic with the solitons passing through each other with practically no deformation. A possible way of preparing the dark-in-bright  soliton by phase imprinting is illustrated. In addition to an isolated dark-in-bright soliton, we also demonstrate the viability of preparing these solitons in a binary dipolar BEC as two coupled 
dark-in-bright solitons or as a bright soliton coupled to a dark-in-bright soliton. The numerical simulation was done by explicitly solving 
the 3D GP equation with realistic values of contact and  dipolar interactions of $^{164}$Dy and $^{162}$Dy  atoms.  The results and conclusions  of the present paper can be  tested in experiments with present-day know-how and 
technology  and shuld lead to interesting future investigations.

%\acknowledgments
We thank  
FAPESP  and  CNPq (Brazil)  for partial support.

\end{document}